\documentclass[proceedings, preprint]{rmaa}

\usepackage{amsmath}
\usepackage{hyperref}
\usepackage{paralist}

\usepackage{psfrag,color}

\SetYear{2013}
\SetConfTitle{Magnetic Fields in the Universe IV}

\title{Equivalence between
formulations in Cosmological Perturbation Theory: The primordial magnetic
fields as an  example.} 

\author{ H\'ector J. Hort\'ua \altaffilmark{1} 
  and Leonardo Casta\~{n}eda, \altaffilmark{1}}

\altaffiltext{1}{Grupo de Gravitaci\'on y Cosmolog\'ia, Observatorio Astron\'omico Nacional,
Universidad Nacional de Colombia, Cra 45 \#  26-85, Ed. Uriel Gutierr\'ez, Bogot\'a D.C, Colombia.}

\shortauthor{Hort\'ua  \& Casta\~{n}eda}
\shorttitle{RevMexAA(SC) Demo Document}

\listofauthors{ H\'ector J. Hort\'ua  \& Leonardo Casta\~{n}eda}
\indexauthor{Hort\'ua, H. J.}
\indexauthor{Casta\~{n}eda, L.}

\abstract{Nowdays, Cosmological Perturbation Theory is a standard and useful tool in theoretical cosmology.  In this work, we compare the 1+3 covariant formalism in perturbation theory (Ellis et al.)  to the gauge invariant approach (Bruni et al.), and we show the equivalence of these formalisms to fix the choice of the perturbed variables (gauge choice) in magnetogenesis. We analyze the evolution of primordial magnetic fields through perturbation theory and we discuss the similarities and differences between these two approaches. We get the Maxwell's equations  and  show a cosmic dynamo like equation written in Poisson gauge, computing the evolution of primordial magnetic fields. Finally, prospects  around these formalisms in the study of magnetogenesis are discussed.}

\resumen{La Teor\'ia de Perturbaciones Cosmol\'ogicas es actualmente  una herramienta est\'andar y \'util en la cosmolog\'ia te\'orica. En este
trabajo se compara el formalismo covariante 1+3  de  la teor\'ia de perturbaciones (Ellis et al.) con la approximaci\'on  invariante
gauge  (Bruni et al.), y mostramos  la equivalencia de estos formalismos al  fijar la escogencia de las 
variables perturbadas  v\'ia  magnetog\'enesis.  Se analiza la evoluci\'on de los campos magn\'eticos primordiales a trav\'es
de la teor\'ia de perturbaciones y se discuten las similitudes y diferencias entre estos dos enfoques. Tambi\'en se muestra
las ecuaciones de Maxwell y la ecuaci\'on tipo  dinamo c\'osmico  escrita en el gauge Poisson el cual da cuenta de la evoluci\'on
de los campos magn\'eticos primordiales. Por \'ultimo, las perspectivas en torno a estos formalismos en el estudio de magnetogenesis son
discutidos.}

\addkeyword{primordial magnetic fields}
\addkeyword{Cosmological Perturbation Theory}
\begin{document}
% Typeset article header
\maketitle
\section{Introduction}
The origin of large scale magnetic fields  is nowdays one of the major unsolved mysteries in cosmology. These fields are assumed to be increased and maintained by dynamo mechanism, but  it needs a \textit{seed} for the mechanism takes place \citep{astro-ph/0207240}. Astrophysical mechanisms,  as the Biermann battery   have been used to explain how the magnetic field is mantained  in  objects  as galaxies, stars and supernova remnants, but they are not likely  correlated  beyond galactic sizes making  difficult to use astrophysical mechanisms to explain the origin of magnetic fields on cosmological scales \citep{0707.2783}. In order to overcome this problem, the primordial origin should be found in other scenarios from  which the astrophysical mechanism start.   For example, magnetic fields could be ge\-nerated during primordial phase transitions (such as QCD, the electroweak or GUT), parity-violating processes which generates magnetic helicity or during inflation \citep{2001}.  The advantage of these primordial processes is that they offer a wide range of coherence lengths many of which are strongly constrained by Nucleosynthesis, while the astrophysical mechanisms  produce fields at the same order of the   astrophysical size of the  object \citep{0907.0197}.   
One way for describe the evolution of magnetic fields is through Cosmological Perturbation Theory. This theory is a  powerful  tool to understand the present properties of the large-scale structure of the Universe and their origin (for a review, see \citep{0201405}). 
The main goal in this paper is to study the late evolution  of magnetic fields  which were generated in early stages of the universe. We use the cosmological perturbation theory following the Gauge Invariant  formalism to find  the perturbed Maxwell equations and  also we obtain a dynamo like equation written in terms of gauge invariant variables. Futhermore,  we discuss the importance that both curvature and the gravitational potential play in the evolution of these fields. The paper is organized as follows: in the next section \S~\ref{gaugeproblem} we briefly give an introduction of cosmological perturbation theory and  we address the gauge problem  in this theory. In the section \S~\ref{sec1} we  consider  the perturbed FLRW and we found the conservation equations at first order written in terms of gauge invariant quantities and the section \S~\ref{2} a magnetic field in the FLRW background is introduced.  Also a discussion  between   1+3 covariant  \citep{1804}  to the gauge invariant \citep{1997} approaches is done in \S~\ref{2}. The final section \S~\ref{error} is devoted to discuss  the main results  and the connection with future works.
\section{
The gauge problem in perturbation theory}
\label{gaugeproblem}
\begin{figure}[!t]
  \includegraphics[width=\columnwidth]{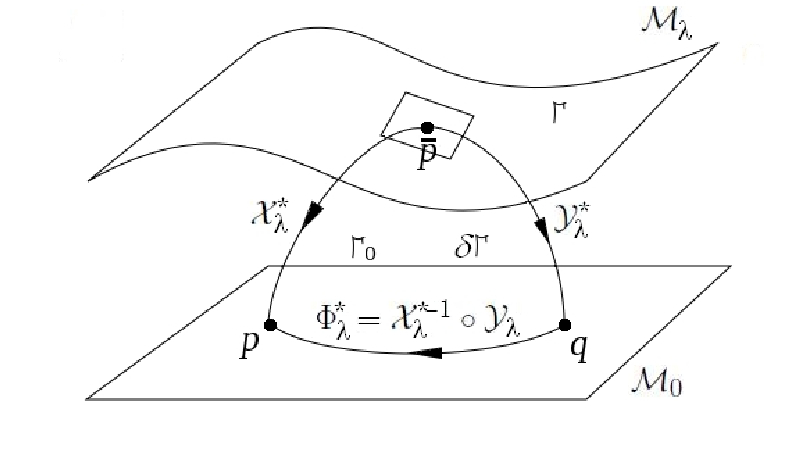}
  \caption{Gauge Transformation.}
  \label{hector}
\end{figure}

Cosmological perturbation theory help us to find approximated solutions of the Einstein field equations through small desviations from  an exact solution. The gauge invariant  formalism is developed into two space-times, one is the real space-time  $(\mathcal{M},g_{\alpha\beta} )$  which  describes the perturbed universe and the other one is the background space-time $(\mathcal{M}_{0},g_{\alpha\beta}^{(0)})$  which is an idealization and  is taken as reference  to generate the real space-time.
 A mapping between these space-times  called \textit{gauge choice} given by a function   $\mathcal{X}:\:\mathcal{M}_{0}(p)\longrightarrow\mathcal{M}(\bar{p})$ for any  point $p\in \mathcal{M}_{0}$ and $\bar{p}\in \mathcal{M}$, which generates a pull-back
\begin{equation}
\begin{array}{c}
\mathcal{X}^{*}:\\
\\\end{array}\;\begin{array}{c}
\mathcal{M}\\
T^{*}(\overline{p})\end{array}\longrightarrow\begin{array}{c}
\mathcal{M}_{0}\\
T^{*}(p)\end{array},\end{equation}
thus, points on the real and background space-time can be compared through of $\mathcal{X}$.
General covariance states that there is no preferred
coordinate system in nature and it introduce a  gauge
in perturbation theory. This  gauge
is an unphysical degree of freedom and we have to fix the  gauge
or to extract some invariant quantities to have physical results \citep{0804}. 
Then, the  perturbation for $\varGamma$ is defined as
\begin{equation}
  \delta\varGamma(p)=\varGamma(\bar{p})-\varGamma^{(0)}(p)\label{perturbation}.
\end{equation}
We see that the perturbation $\delta\varGamma$ is completely dependent of the gauge choice because the mapping determines the representation on $\mathcal{M}_{0}$ of  $\varGamma(\bar{p})$.   However, one can also choice another correspondence $\mathcal{Y}$ between these space-times (see Figure \ref{hector}) so that $\mathcal{Y}:\,\mathcal{M}_{0}(q)\rightarrow\mathcal{M}(\overline{p})$, ($p\neq q$).   The  freedom  to choose  different correspondences  generate an  arbitrariness in the value of $\delta\varGamma$ at any space-time point $p$, which is called  \textit{gauge problem} \citep{1882}.  Given a tensor field $\varGamma$ , the relations between
first and second order perturbations of $\varGamma$ in two different gauges are
\begin{align}
\delta^{(1)}_{\mathcal{Y}}\varGamma-\delta^{(1)}_{\mathcal{X}}\varGamma & =\mathcal{L}_{\xi_{1}}\varGamma_{0},\label{eq1}\\
\delta^{(2)}_{\mathcal{Y}}\varGamma-\delta^{(2)}_{\mathcal{X}}\varGamma & = 2\mathcal{L}_{\xi_{1}}\delta^{(1)}_{\mathcal{X}}\varGamma_{0}+\left(\mathcal{L}_{\xi_{1}}^{2}+\mathcal{L}_{\xi_{2}}\right)\varGamma_{0},\label{2transform}\end{align}
where the generators of the gauge transformation $\Phi$ are
\begin{equation}
\xi_{1}=Y-X  \quad \text{and} \quad \xi_{2}=\left[ X,Y\right].
\end{equation}
This vector field can be split in their time and space part
\begin{equation}
 \xi^{(r)}_{\mu}\rightarrow\left(\alpha^{(r)},\partial_{i}\beta^{(r)}+d_{i}^{(r)} \right) ,
\end{equation}
here $\alpha^{(r)}$ and $\beta^{(r)}$ are arbitrary scalar functions, and $\partial^{i}d_{i}^{(r)}=0$. The function $\alpha_{(r)}$ determines the choice of constant time hypersurfaces, while $\partial_{i}\beta^{(r)}$ and $d_{i}^{(r)}$  fix the spatial coordinates within these hypersurfaces. The choice of coordinates is arbitrary  and the definitions of perturbations are thus gauge dependent. 
The gauge transformation given by the equations (\ref{eq1}) and (\ref{2transform}) are quite general. To first order  $\varGamma$ is gauge invariant if  $\mathcal{L}_{\xi_{1}}\varGamma_{0}=0$, while to second order one must have another conditions  $\mathcal{L}_{\xi_{1}}\delta^{(1)}_{\mathcal{X}}\varGamma_{0}=\mathcal{L}_{\xi_{1}}^{2}\varGamma_{0}=0$ and $\mathcal{L}_{\xi_{2}}\varGamma_{0}=0$, and so on at high orders \citep{a471}.
We will apply the formalism described above to the Robertson-Walker metric, where $k$ does mention  the expansion order. 
\section{Gauge invariant variables at first order}
\label{sec1}
We consider the  perturbations about a FLRW background, so  the metric tensor is given by :
\begin{align}
g_{00}&=-a^{2}(\tau)\left(1+2\sum_{k=1}^{\infty}\frac{1}{k!}\psi^{(k)}\right),\\
g_{0i}&=a^{2}(\tau)\sum_{k=1}^{\infty}\frac{1}{k!}\omega_{i}^{(k)},\\
g_{ij}&=a^{2}(\tau)\left[\left(1-2\sum_{k=1}^{\infty}\frac{1}{k!}\phi^{(k)}\right)\delta_{ij}\right.\left.+\sum_{k=1}^{\infty}\frac{\chi_{ij}^{(k)}}{k!}\right]. \end{align}
The perturbations are splitting into scalar, transverse vector part, and transverse trace-free tensor
\begin{equation}
\omega_{i}^{(k)}=\partial_{i}\omega^{(k)\Vert}+\omega_{i}^{(k)\bot}, \label{omega1}\end{equation}
with $\partial^{i}\omega_{i}^{(k)\bot}=0$. Similarly we can split $\chi_{ij}^{(k)}$ as
\begin{equation}
\chi_{ij}^{(k)}=D_{ij}\chi^{(k)\Vert}+\partial_{i}\chi_{j}^{(k)\bot}+\partial_{j}\chi_{i}^{(k)\bot}+\chi_{ij}^{(k)\top}, \label{chi1}
\end{equation}
for any tensor quantity.\footnote{With $\partial^{i}\chi_{ij}^{(k)\top}=0$, $\chi_{i}^{(k)i}=0$  and $D_{ij}\equiv\partial_{i}\partial_{j}-\frac{1}{3}\delta_{ij}\partial_{k}\partial^{k}$.} 
Following, one can find the scalar \textit{gauge invariant variables} at first order given by
\begin{eqnarray}
\Psi^{(1)}&\equiv&\psi^{(1)}+\frac{1}{a}\left(\mathcal{S}_{(1)}^{||}a\right)^{\prime},\\
\Phi^{(1)}&\equiv&\phi^{(1)}+\frac{1}{6}\nabla^{2}\chi^{(1)}-H\mathcal{S}_{(1)}^{||},\\
\Delta^{(1)}&\equiv&\mu_{(1)}+\left(\mu_{(0)}\right)^{\prime}\mathcal{S}_{(1)}^{||},\\
\Delta^{(1)}_{P}&\equiv& P_{(1)}+\left(P_{(0)}\right)^{\prime}\mathcal{S}_{(1)}^{||},
\end{eqnarray}
with $\mathcal{S}_{(1)}^{||}\equiv\left(\omega^{||(1)}-\frac{\left(\chi^{||(1)}\right)^{\prime}}{2}\right)$ the scalar contribution of the shear. The vector modes are
\begin{eqnarray}
{\upsilon}_{(1)}^{i}&\equiv&v_{(1)}^{i}+\left(\chi_{\bot(1)}^{i}\right)^{\prime},\\
\vartheta_{i}^{(1)}&\equiv&\omega_{i}^{(1)}-\left(\chi_{i}^{\bot(1)}\right)^{\prime},\\
\mathcal{V}_{(1)}^{i}&\equiv&\omega_{(1)}^{i}+v_{(1)}^{i}.
\end{eqnarray}
\begin{equation}
\mu=\mu_{0}+\sum_{r=1}^{\infty}\frac{1}{r!}\delta^{r}\mu, \,  u^{\alpha}=\frac{1}{a}\left(\delta_{0}^{\alpha}+\sum_{r=1}^{\infty}\frac{1}{r!}v_{(r)}^{\alpha}\right)  \end{equation}
with $\mathcal{S}_{(1)}^{||}\equiv\left(\omega^{||(1)}-\frac{\left(\chi^{||(1)}\right)^{\prime}}{2}\right)$ the scalar contribution of the shear (associated with $\alpha^{(1)}$). Using the Einstein's equation at first order, it is expressed the evolution of geometrical variables $\phi$ and $\psi$, and the conservation's equations entails the evolution of  energy density $\Delta^{(1)}$ 
\begin{eqnarray}
&&\left(\Delta^{(1)}\right)^{\prime}  +3H\left(\Delta_{P}^{(1)}+\Delta^{(1)}\right) \nonumber\\
&-&3\left(\Phi^{(1)}\right)^{\prime}\left(P_{(0)}+\mu_{(0)}\right) +\left(P_{(0)}+\mu_{(0)}\right)\nabla^{2}\upsilon^{(1)}\nonumber \\
&-&3H\left(P_{(0)}+\mu_{(0)}\right)^{\prime}\mathcal{S}_{(1)}^{||}-\left(\left(\mu_{(0)}\right)^{\prime}\mathcal{S}_{(1)}^{||}\right)^{\prime} \nonumber\\
&+&\left(P_{(0)}+\mu_{(0)}\right)\left(-\frac{1}{2}\nabla^{2}\chi^{(1)}+3H\mathcal{S}_{(1)}^{||}\right)^{\prime} \nonumber\\
&-&\left(P_{(0)}+\mu_{(0)}\right)\nabla^{2}\left(\frac{1}{2}\chi^{||(1)}\right)^{\prime}=0, \label{e.c}
\end{eqnarray}
and peculiar velocity $\mathcal{V}_{(1)}$ given by 
\begin{eqnarray}
&&\left(\mathcal{V}_{i}^{(1)}\right)^{\prime}+\frac{\left(\mu_{(0)}+P_{(0)}\right)^{\prime}}{\left(\mu_{(0)}+P_{(0)}\right)}\mathcal{V}_{i}^{(1)}-4H\mathcal{V}_{i}^{(1)}+\partial_{i}\Psi^{(1)} \nonumber\\
&-&\frac{_{\partial_{i}\left(\Delta_{P}^{(1)}-\left(P_{(0)}\right)^{\prime}\mathcal{S}_{(1)}^{||}\right)+\partial_{l}\Pi_{(fl)i}^{(1)l}}}{\left(\mu_{(0)}+P_{(0)}\right)}=\frac{\partial_{i}\left(\mathcal{S}_{(1)}^{||}a\right)^{\prime}}{a}.\label{e.ne}\end{eqnarray}
\section{Specifying to Poisson gauge}
It is possible to fix the four degrees of freedom by imposing gauge conditions.
 If we impose  the gauge restrictions given in \citep{ber} we can removing the  degrees of freedom,  we fix the gauge conditions as
 \begin{equation}
 \partial^{i}\omega_{i}^{(r)}=\partial^{i}\chi_{ij}^{(r)}=0,
\end{equation}
this lead to some functions are dropped
 \begin{equation}
 \omega^{\parallel(r)}=\chi^{(r)\bot}_{i}=\chi^{(r)\parallel}=0,
\end{equation}
with the functions defined in equations (\ref{omega1}) and (\ref{chi1}). Using the last constraints  together with equations (5.18)-(5.21) in \citep{1997} and following the procedure made in \citep{malik}, the vector which determines the gauge transformation at first order $\xi_{i}^{(1)}=\left(\alpha^{(1)},\partial_{i}\beta^{(1)}+d^{(1)}_{i}\right)$ is given by,
\begin{equation}
\alpha^{(1)}  \rightarrow  \omega_{(1)}^{\parallel}+\beta^{\prime}_{(1)},\quad \beta^{(1)}  \rightarrow -\frac{\chi^{\parallel(1)}}{2},\quad 
d_{i}^{(1)}  \rightarrow -\chi^{\bot(1)}_{i}. \label{apen1}
\end{equation}
\section{Weakly magnetized FLRW-background}
\label{2}
 We   allow the presence of a weak magnetic field into our FLRW space-time with the  property $B_{(0)}^{2}\ll\mu_{(0)}$ and must  be sufficiently random to satisface $\left\langle B_{i} \right\rangle=0$ and $ \left\langle B_{(0)}^{2}\right\rangle \neq0$ to ensure that  symmetries and the evolution of the background  remain unaffected 
(we assume that at zero order the  magnetic field has been generated by some random process which is statistically
homogeneous so that  $B_{(0)}^{2}$ just time depending and  $\left\langle .. \right\rangle$ denotes the expectation value) \citep{barrow}.
We work under MHD approximation, thus,  in large scales the plasma is globally neutral, charge density is neglected and the electric field with the current  should be zero, thus the only zero order magnetic variable is $B_{(0)}^{2}$. At first order it is obtained a gauge invariant term which describes the magnetic energy density
\begin{equation}
\Delta^{(1)}_{mag}\equiv\ B_{(1)}^{2}+\left(B_{(0)}^{2}\right)^{\prime}\alpha^{(1)} \label{binvgauge}.\end{equation}
Another gauge invariant variables are  the 3-current $J$,  the charge density $\varrho$ and the electric and magnetic fields, because they vanish in the background.\\ 
At first order, the electric field and the current become nonzero  and assuming the ohmic current is not neglected, we find the Ohm's law  
\begin{equation}
J_{i}^{(1)}=\sigma\left[  E_{i}^{(1)}+\left(\mathcal{V}^{(1)}\times B^{(0)}\right)_{i}\right]. \label{11ohm}\end{equation}
The perturbed equations for the metric and electromagnetic fields are given by
\begin{eqnarray}
&&\partial_{i}E_{(1)}^{i}=a\varrho_{(1)},\quad \epsilon^{ilk}\partial_{l}B^{(1)}_{k}=\left(a^{2}E_{(1)}^{i}\right)^{\prime}+a^{3}J_{(1)}^{i},\label{3ME} \nonumber\\
&&\partial^{i}B_{i}^{(1)}=0,\quad \left(a^{2}B_{k}^{(1)}\right)^{\prime}+a^{2}\epsilon_{\: k}^{ij}\partial_{i}E_{j}^{(1)}=0.\label{12}\end{eqnarray}
Now using equation (\ref{12}) together with the Ohm's law, we get a cosmic dynamo like equation which describes  the evolution of density magnetic field at first order in the Poisson gauge
\begin{eqnarray}
&&\frac{d\Delta_{(mag)}^{(2)}}{dt}+4H\Delta_{(mag)}^{(2)}+2\eta \mathcal{\epsilon}^{(2)} 
\nonumber \\
&&+\frac{2}{3} \Delta_{(mag)}^{(1)}\partial_{l}\mathcal{V}_{(1)}^{l}=-2\Pi^{(1)}_{ij(em)}\sigma^{ij}_{(1)}\label{dyn}, \end{eqnarray}
with $\mathcal{\epsilon}^{(2)}$ given by
\begin{eqnarray}
&&\mathcal{\epsilon}^{(2)}=-B^{(1)} \cdot \nabla^{2}B^{(1)} \nonumber \\
&&- B^{(1)}_{k} \left(\nabla \times \left(\frac{dE_{(1)}}{dt}+2HE_{(1)}\right)\right)^{k} \nonumber\\
&& -\frac{\Delta_{(mag)}^{(1)}}{2}\nabla^{2}\left( \Psi^{(1)}-3 \Phi^{(1)} \right) \nonumber\\
&& +B^{k}_{(1)}\left(B^{(0)}\cdot \nabla \right)\partial_{k}\left( \Psi^{(1)}-3 \Phi^{(1)}\right), \end{eqnarray}
%\begin{eqnarray}
%&&\frac{d\Delta_{(mag)}^{(2)}}{dt} +4H\Delta_{(mag)}^{(2)}+2\eta \left[-B^{(1)} \cdot \nabla^{2}B^{(1)}  \right. \nonumber \\
%&&- B^{(1)}_{k} \cdot \left(\nabla \times \left(\frac{dE_{(1)}}{dt}+2HE_{(1)}\right)\right)^{k} \nonumber \\
%&& \left. -\frac{\Delta_{(mag)}^{(1)}}{2}\nabla^{2}\left( \Psi^{(1)}-3 \Phi^{(1)} \right)+B^{k}_{(1)}\left(B^{(0)}\cdot \nabla \right)\partial_{k}\left( \Psi^{(1)}-3 \Phi^{(1)}\right) \right] +\frac{2}{3} \Delta_{(mag)}^{(1)}\partial_{l}\mathcal{V}_{(1)}^{l}=-2\Pi^{(1)}_{ij(em)}\sigma^{ij}_{(1)}\label{dyn}, \end{eqnarray}
where we use the Lagrangian coordinates which are comoving with the local Hubble  flow and magnetic field lines are frozen into the fluid ($\dfrac{d}{dt}=\dfrac{\partial}{\partial t}+\mathcal{V}_{(1)}^{i}\partial_{i} $), $\sigma$ and $\Pi$ are the shear and stress Maxwell tensor respectively. 
Thus, the perturbations in the space-time play an important role in the evolution of primordial magnetic fields \citep{hortua}.  In the case of a homogeneous collapse, $B \sim \mathcal{V}^{-\frac{2}{3}}$ there is  an amplification of  the magnetic field in places where gravitational collapse take place. In equation (\ref{dyn}), the energy density magnetic field at second order transforms as
\begin{eqnarray}
&& \Delta_{(mag)}^{(2)}= B^{2}_{(2)}+\alpha_{(1)}\left(B^{2 \prime \prime}_{(0)}\alpha_{(1)}+B^{2 \prime}_{(0)}\alpha_{(1)}^{\prime}+2B^{2 \prime}_{(1)} \right)  \nonumber\\
&&+  \xi^{i}_{(1)}\left(B^{2\prime}_{(0)}\partial_{i}\alpha^{(1)}+2\partial_{i}B^{2}_{(1)}\right) +B^{2 \prime}_{(0)} \alpha_{(2)}.
 \end{eqnarray}
Finally, we  relate quantities in the 1+3 covariant formalism and in the
invariant approach showed above.  
In the covariant formalism quantities are projected down
onto spatial $h_{\alpha\beta}$, relative to the  4-velocity of the fluid. With this choice, we expect that covariant formalism could be equivalent to  comoving gauge \citep{comovil}. The comoving  magnetic density
gradient  is defined as
\begin{equation}
{\cal B}_\mu\equiv\frac{a}{B^{2}}h^\lambda_\mu\nabla_\lambda B^{2}, \quad \text{with}\quad h_{\mu\nu}\equiv g_{\mu\nu}+u_\mu u_\nu\,. 
 \end{equation} Now, following \citep{2012} we substitute  the 4-velocity at first order found in gauge invariant approach $u_\mu=a\left(  -(1+\psi),\ \partial_{i}(v^{\parallel}+\omega^{\parallel})\right) \,$,    obtaining  the following relation
\begin{equation}
h^\lambda_\mu\nabla_\lambda B^{2}=\partial_{i}\left( B^{2}_{(1)}+B^{2 \prime}_{(0)}\left(v^{\parallel}+\omega^{\parallel}\right)\right) \,.
\label{a} \end{equation}  If the comoving gauge is used (which introduces a family of world lines orthogonal to the 3-D spatial sections) given by $\alpha \rightarrow v^{\parallel}+\omega^{\parallel}$ in equation  (\ref{binvgauge}),   is derived a similar to expression as it was found in 1+3 covariant formalism in  equation  (\ref{a}),  implying an equivalence in both formalisms at linear order.
Now, if we study this equivalence at second order, we must impose  $u_{i}^{(2)}=0$ to provide a covariant description (due to we must ensure 4-velocity orthogonal to 3D spatial sections). In this case the 4-velocity at second order is
\begin{equation}
\frac{ u_{i}^{(2)}}{a}=\left[\frac{(v_{i}^{(2)}+\omega_{i}^{(2)})}{2}-2v_{i}^{(1)}\phi^{(1)}-\omega^{(1)}_{i}\psi_{(1)}+v^{j}_{(1)}\chi_{ij}^{(1)}\right]\,\end{equation}
in \citep{tesis}  is shown  vector field  that determines the gauge comoving at second order. In this case one must take into account that 4-velocity must be zero and choose appropriately the 3D spatial section through of $\beta^{(2)}$ and $d_{i}^{(2)}$. 
\section{discussion}
\label{error}
A problem in modern cosmology is to explain the origin of cosmic magnetic fields. The origin of these fields is still in debate but  they must affect  the formation of large scale structure and the anisotropies in the cosmic microwave background radiation (CMB) \citep{2008}. 
We observe that essentially, the functional form are the same in the two approaches, the coupling between geometrical perturbations and fields variables appear as sources in the magnetic field evolution giving a new possibility to explain the amplification of primordial cosmic magnetic fields.  Thus the perturbations in the space-time play an important role in the evolution of primordial magnetic fields. The equation  (\ref{dyn})  is  dependent  on geometrical quantities (perturbation in the gravitational potential, curvature, velocity ...).  These quantities evolve  according to the Einstein field equations (the Einstein field equation to second order are given in \citep{2007}). In this way, the equation equation  (\ref{dyn}) tell us how the magnetic field evolves according to scale of the perturbation. In  sub-horizon scale, the contrast density and the geometrical quantities grow. Hence, is expected that naturally  the dynamo term should amplify the magnetic field.

\end{document}